\documentclass{aastex}          
\usepackage{spr-astr-addons}    
\usepackage{url}
\usepackage{graphicx}	
\usepackage{amsmath}	
\usepackage{amssymb}	
\usepackage{tabto}
\usepackage{makecell}
\usepackage{booktabs}
\usepackage{rotating}
\usepackage[caption=false]{subfig}

\newcommand{\ngc}{J054149.24--641813.7}

\newcommand{\ergcm}[1]{erg\,cm$^{-2}$\,s$^{-1}$}
\def\HI{\hbox{H{\sc i}}}
\def\HII{\hbox{H{\sc ii}}}

\newcommand{\emaila}{jbalzan.astro@gmail.com}
\newcommand{\emailb}{M.Filipovic@westernsydney.edu.au}
\newcommand{\emailc}{Shi.Dai@westernsydney.edu.au}
\newcommand{\emaild}{19158264@student.westernsydney.edu.au}
\newcommand{\emaile}{l.barnes@westernsydney.edu.au}

\begin{document}
%
\title{A Radio Continuum Study of NGC\,2082}

\shorttitle{Radio Continuum Study: NGC\,2082}
\shortauthors{Balzan et al.}

\author{Balzan J. C. F. \altaffilmark{1}\\ \emaila}
\and \author{Filipovi\'c M. D. \altaffilmark{1}\\ \emailb}
\and \author{Dai S. \altaffilmark{1,2}\\ \emailc}
\and \author{Alsaberi R. Z. E. \altaffilmark{1}\\ \emaild}
\and \author{Barnes L. \altaffilmark{1}\\ \emaile}

\affil{$^1$Western Sydney University, Locked Bag 1797, Penrith South DC, NSW 1797, Australia}
\affil{$^2$CSIRO Astronomy and Space Sciences, Australia Telescope National Facility, PO Box 76, Epping, NSW 1710, Australia}


\begin{abstract}
We present radio continuum observations of NGC\,2082 using ASKAP, ATCA and Parkes telescopes from 888\,MHz to 9000\,MHz. Some 20~arcsec from the centre of this nearby spiral galaxy, we discovered a bright and compact radio source, \ngc{}, of unknown origin. To constrain the nature of \ngc{}, we searched for transient events with the Ultra-Wideband Low Parkes receiver, and compare its luminosity and spectral index to various nearby supernova remnants (SNRs), and fast radio burst (FRB) local environments. Its radio spectral index is flat ($\alpha = 0.02 \pm 0.09$) which is unlikely to be either an SNR or pulsar. No transient events were detected with the Parkes telescope over three days of observations, and our calculations show \ngc{} is two orders of magnitude less luminous than the persistent radio sources associated with FRB\,121102 \& 190520B. We find that the probability of finding such a source behind NGC\,2082 is $P = 1.2\%$, and conclude that the most likely origin for \ngc{} is a background quasar or radio galaxy. 
\end{abstract}

\keywords{Radio continuum: galaxies, Galaxies: spiral, Galaxies: ISM}

\section{Introduction}
\label{intro}

In the absence of an active galactic nucleus (AGN), a spiral galaxy's radio emission primarily derives from non-thermal synchrotron radiation from supernova remnants (SNRs), and thermal bremsstrahlung from \HII\ regions \citep{1992ARA&A..30..575C, book1, book2}. Thus, deep and wide radio surveys from the new generation of radio telescopes such as the Australian Square Kilometre Array Pathfinder (ASKAP) and MeerKAT can shed important light on the processes by which star formation shapes the interstellar medium (ISM). 

NGC\,2082 is G-type spiral galaxy (of SB(r)b morphology) in the Dorado constellation. It has an absolute B-band magnitude of $M_B = 12.79$ \citep{appmag}, a diameter of 10.16\,kpc, and is located at a distance of 18.5\,Mpc \citep{NGC2082distance} and redshift $z = 0.00395$. Unlike some other galaxies in the Dorado constellation (e.g. NGC\,1566), NGC\,2082 remains poorly studied, with its most notable feature being a type~II supernova, SN1992ba \citep{SN1992ba}.

Here, we study the radio properties of NGC\,2082 using ASKAP, Australia Telescope Compact Array (ATCA) and Parkes radio telescope observations. We will also draw on Hubble Space Telescope (HST) observations. Section~\ref{observations} presents our observations and data analysis of NGC\,2082. Section~\ref{results} gives our results and discussion, and conclusions are presented in Section~\ref{conclusion}.

\section{Observations \& Data}
\label{observations}
NGC\,2082 has been observed in the ASKAP--EMU 888\,MHz radio continuum survey of the Large Magellanic Cloud \citep[LMC;][]{ASKAPLMC,2022MNRAS.512..265F}, as well as in the ATCA 20\,cm mosaic survey \citep{ATCA20}. We have also made new observations from Parkes radio telescope, and obtained new and archival data from ATCA (pre-CABB) and the HST.

\subsection{Australian Telescope Compact Array}
\subsubsection{CABB}
We observed NGC\,2082 on 2019 November 30$^\mathrm{th}$ using the Australian Telescope Compact Array (ATCA) (project code C3275 with 1.5C array configuration). The observations were carried out in `snap-shot' mode, with 1-hour of integration over a 12 hour period as a minimum. We used the Compact Array Broadband Backend (CABB) (with 2048\,MHz bandwidth), centred at a) wavelengths of 3/6\,cm ($\nu$~=~4500--6500 and 8000--10000\,MHz), totalling 43.2\,minutes of integration and, b) 13\,cm ($\nu$~=~2100\,MHz), totalling 43.2\,minutes of integration. The primary calibrator (flux) was PKS\,B1934--638 and secondary calibrator (phase) was PKS\,B0530--727. 

The \textsc{miriad}\footnote{\url{http://www.atnf.csiro.au/computing/software/miriad/}} \citep{1995ASPC...77..433S} and \textsc{karma}\footnote{\url{http://www.atnf.csiro.au/computing/software/karma/}} \citep{1995ASPC...77..144G} software packages were used for reduction and analysis. Imaging was completed using the multi-frequency synthesis \textsc{invert} task with natural Briggs weighting (robust=0 for all images), and beam size of $4.5\times4.1$\,arcsec, $1.9\times1.8$\,arcsec, and $1.3\times1.0$\,arcsec for 2100, 5500, and 9000\,MHz images, respectively. The \textsc{mfclean} and \textsc{restor} algorithms were used to deconvolve the images, with primary beam correction applied using the \textsc{linmos} task. We follow the same process with Stokes {\it Q} and {\it U} parameters to produce polarisation maps, except with beam size of $5\times5$\,arcsec (see Section~\ref{sec:pol} below).

\subsubsection{Pre-CABB}
We analysed archival\footnote{Australia Telescope Online Archive (ATOA), hosted by the Australia Telescope National Facility (ATNF): \url{https://atoa.atnf.csiro.au}} ATCA data (project code C466 with 6A array configuration) from 16$^{\rm th}$ November 1995. The observations used pre-CABB receiver (with 128\,MHz bandwidth), centered at wavelengths of 3/6\,cm (4800 and 8640\,MHz) totalling of 48\,minutes of integration. The primary calibrator (flux) was PKS\,B1934--638 and the secondary calibrator (phase) was PKS\,B0355-483.

\subsection{Australian Square Kilometre Array Pathfinder}
NGC\,2082 was observed serendipitously in the ASKAP-EMU radio continuum survey of the Large Magellanic Cloud \citep{ASKAPLMC}, at the edge of the 120\,deg$^{2}$ field. This survey was performed at 888\,MHz with a 288\,MHz bandwidth and $13.9\times12.1$~arcsec beam size.

\subsection{Parkes Radio Telescope}\label{subsec:Parkes}
We observed NGC\,2082 on 2021 June 14$^\mathrm{th}$, 15$^\mathrm{th}$ and 18$^\mathrm{th}$ with the Parkes radio telescope (project code PX075) using the Ultra-Wideband-Low (UWL) receiver \citep{uwl}, which delivers radio frequency coverage from 704\,MHz to 4032\,MHz. All observations were pointed at \ngc{} and executed using the transient search mode where data are recorded with 2-bit sampling every 64\,$\mu$s in each of the 0.125\,MHz wide frequency channels (26624 channels across the whole band).

The full UWL band was split into multiple 512\,MHz subbands for the search of bursts. The search was performed using the pulsar searching software package {\sc PRESTO}~\citep{2001PhDT.......123R}. Radio-frequency interference (RFI) were identified and marked using the {\sc PRESTO} routine {\sc RFIFIND} with a 1\,s integration time. To determine the optimal dispersion measure (DM) steps of the search, we used the {\sc DDPLAN.PY} routine of {\sc PRESTO} for a DM range of 200 to 3000\,cm$^{-3}$\,pc. Data were then dedispersed at each of the trial DMs using the {\sc PREPDATA} routine with RFI removal based on the mask file produced by {\sc RFIFIND}. 

Single pulse candidates with a signal-to-noise ratio larger than seven were identified using the {\sc SINGLE\_PULSE\_SEARCH.PY} routine for each dedispersed time series and for different boxcar filtering parameters (from 1 to 300 samples). Burst candidates were manually examined.

\subsection{Hubble Space Telescope}
\begin{figure*}[ht!]
\centering
 \includegraphics[width=17cm]{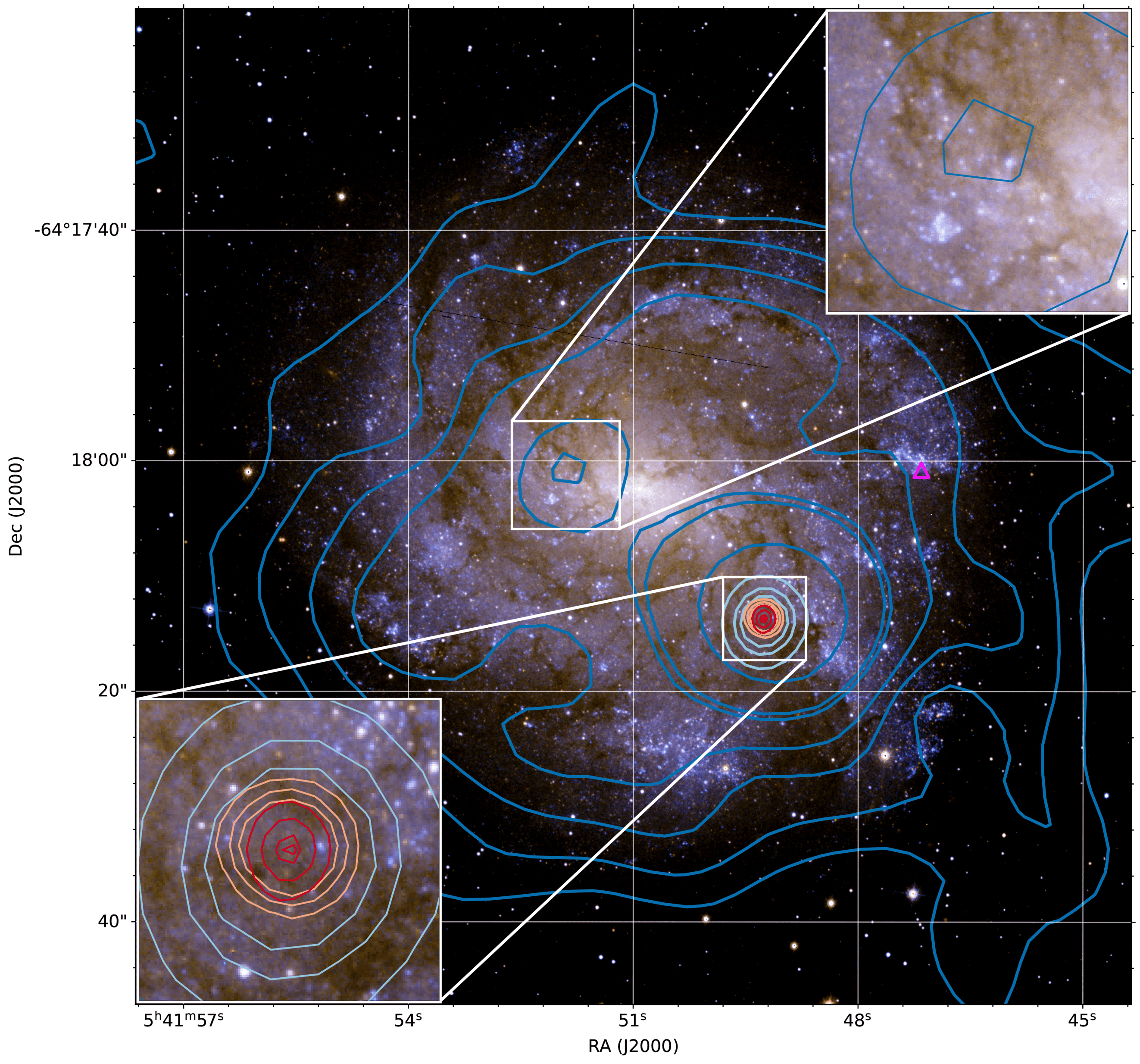}
 \caption{3-colour HST image of NGC\,2082 overlaid with ASKAP and ATCA contours. The blue contours are ASKAP 888\,MHz (0.3, 0.5, 0.7, 0.9, 1.3, 1.4, 2.1\,mJy~beam$^{-1}$), and the light-blue, orange and red contours are ATCA 2100~MHz (0.4, 1.0, 1.5\,mJy~beam$^{-1}$), 5500~MHz (0.5, 1.0, 1.5\,mJy~beam$^{-1}$), and 9000\,MHz (0.4346, 1.5198, 3.0, 3.4\,mJy~beam$^{-1}$) respectively. The inset image in the bottom-left provides a zoom-in of \ngc{}, showing the absence of any optical counterpart. The subplot in the top-right provides a zoom-in of a 888\,MHz flux density peak of 1.3\,mJy~beam$^{-1}$, almost directly opposite \ngc{}. The magenta triangle denotes the position of SN1992ba.}
 \label{fig:contours}
\end{figure*}

NGC\,2082 was first imaged by the Hubble Space Telescope in 1997, following the type II supernova SN1992ba \citep{SN1992ba}, revealing a bright, face-on spiral galaxy (Fig.~\ref{fig:contours}). Fig.~\ref{fig:contours} is a 3-colour image created with APLpy \citep{2012ascl.soft08017R}, using archival HST data\footnote{Based on observations made with the NASA/ESA Hubble Space Telescope, and obtained from the Hubble Legacy Archive, which is a collaboration between the Space Telescope Science Institute (STScI/NASA), the Space Telescope European Coordinating Facility (ST-ECF/ESA) and the Canadian Astronomy Data Centre (CADC/NRC/CSA).} \citep{carollo}, where the red channel uses I-band data (F814W filter), the blue channel uses B-band data (F435W filter), and the green channel is pseudo-green which has been constructed by stacking the red and blue channels (B+I-band). No detection of SN1992ba is reported; it cannot be seen in any of these images.

\section{Results and Discussion}
 \label{results}
A striking feature in all our radio images of NGC\,2082 is strong point radio source (\ngc{}) positioned 20~arcsec from the galaxy centre, as seen in the bottom-left image of Fig.~\ref{fig:contours} by the ATCA contours. We also note no detection of SN1992ba in any of our images. The 9000\,MHz ATCA observations, with our highest resolution, shows an unresolved point radio source, regardless of the parameters used in the data reduction. The top-right subplot in Fig.~\ref{fig:contours} provides a better look at an 888\,MHz emission peak flux density of 0.0013\,Jy~beam$^{-1}$ opposing \ngc{}, and also away from the centre of galaxy. It is unlikely that the two radio sources are related. Finally, we note HST observations with F435W and F814W filters show no optical counterparts to either source, and there are no counterparts at other wavelengths. 

\subsection{Spectral index}
 \label{sec:SI}
\begin{figure}[ht!]
    \centering
    \includegraphics[width=\columnwidth]{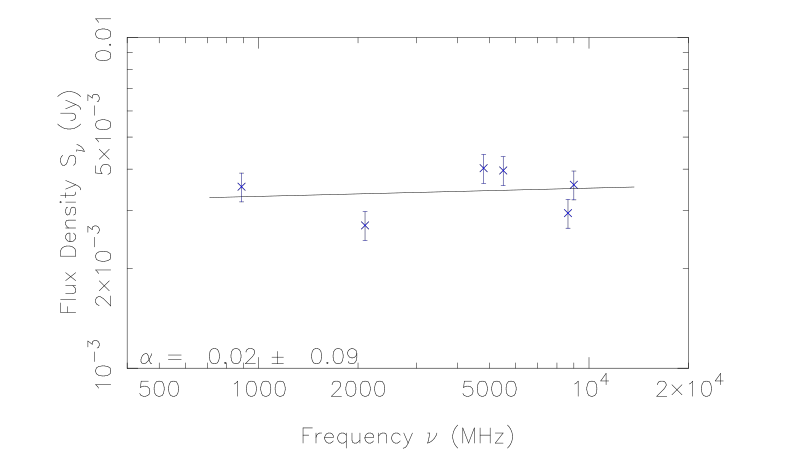}
    \caption{The spectral index of \ngc{} ($\alpha = 0.02 \pm 0.09$) with assumed 10\% error bars, based on the measured flux densities from Table \ref{tab:flux_density}.}
    \label{fig:spectral index}
\end{figure}

In Table~\ref{tab:flux_density} we show the flux densities of \ngc{}, measured using \textsc{CARTA}\footnote{\url{https://cartavis.org/}} and treated as a point source. We assume that the flux density errors are $<$10~per~cent. We estimate a flat radio spectral index of $\alpha = +0.02 \pm 0.09$ suggesting that the emission is predominantly of thermal origin if the source is located in NGC\,2082 (Fig.~\ref{fig:spectral index}). Such a flat spectral index would be very unusual among SNRs and radio pulsar sources \citep{Uro_evi__2014,PSRspecindex,2015MNRAS.449.3223D} unless this source is an unresolved pulsar wind nebulae (PWN). However, a background galaxy (quasar) could explain this radio spectrum (see Section~\ref{sec:bckg}).


\begin{table}
\centering
\caption{The radio flux densities of NGC\,2082's extended radio emission and radio point source \ngc{} (point source). A '---' represents a non-detection. }
\begin{tabular}{lcccccc}
\toprule
Freq. \& Telescope & S$_{\rm NGC\,2082}$ & S$_{\rm point\,source}$   \\
(MHz)     &            (mJy)         & (mJy)    \\
\toprule
888  \& {\footnotesize ASKAP}         & 13.4$\pm$1.3 & 3.5$\pm$0.4 \\
2100 \& {\footnotesize ATCA CABB}     & 11.2$\pm$1.1 & 2.7$\pm$0.3 \\
4800 \& {\footnotesize ATCA Pre-CABB} & --- & 4.0$\pm$0.4 \\ 
5500 \& {\footnotesize ATCA CABB }    & --- & 4.0$\pm$0.4  \\
8640 \& {\footnotesize ATCA Pre-CABB} & --- & 3.0$\pm$0.3 \\ 
9000 \& {\footnotesize ATCA CABB}     & --- & 3.6$\pm$0.4 \\
\bottomrule
$\alpha \pm \Delta \alpha$  &   --0.15$\pm$0.23 & +0.02$\pm$0.09 \\
\bottomrule
\end{tabular}
\label{tab:flux_density}
\end{table}

To measure the flux densities of NGC\,2082's entire extended emission, we use the method described in \cite{hurley-walker} and \cite{2019PASA...36...48H}, which includes careful region selection that also excludes \ngc{}. We measure reliable NGC\,2082 flux densities at two frequencies (888 and 2100~MHz; Table~\ref{tab:flux_density}), which allows us to estimate the spectral index of $\alpha = -0.15 \pm 0.23$. This flat radio spectral index is unusual for spiral galaxies \cite{1982A&A...116..164G}, but consistent with thermal emission from \HII\ regions across NGC\,2082. As the nucleus of NGC\,2082 does not show any radio compact source or emission above 0.1\,mJy~beam$^{-1}$, we suggest this might account for the unusually flat radio spectral index.

\subsection{Polarisation}
 \label{sec:pol}

\begin{figure*}[!ht]
\centering
\subfloat[Fractional polarisation map of \ngc{} at 5500\,MHz, where average $P_{5500\,\text{MHz}} = 6 \pm 2\,\%$. \label{fig:pol_frac_5500}]{\includegraphics[angle=270,width=.45\textwidth, trim=0 20 45 0,clip]{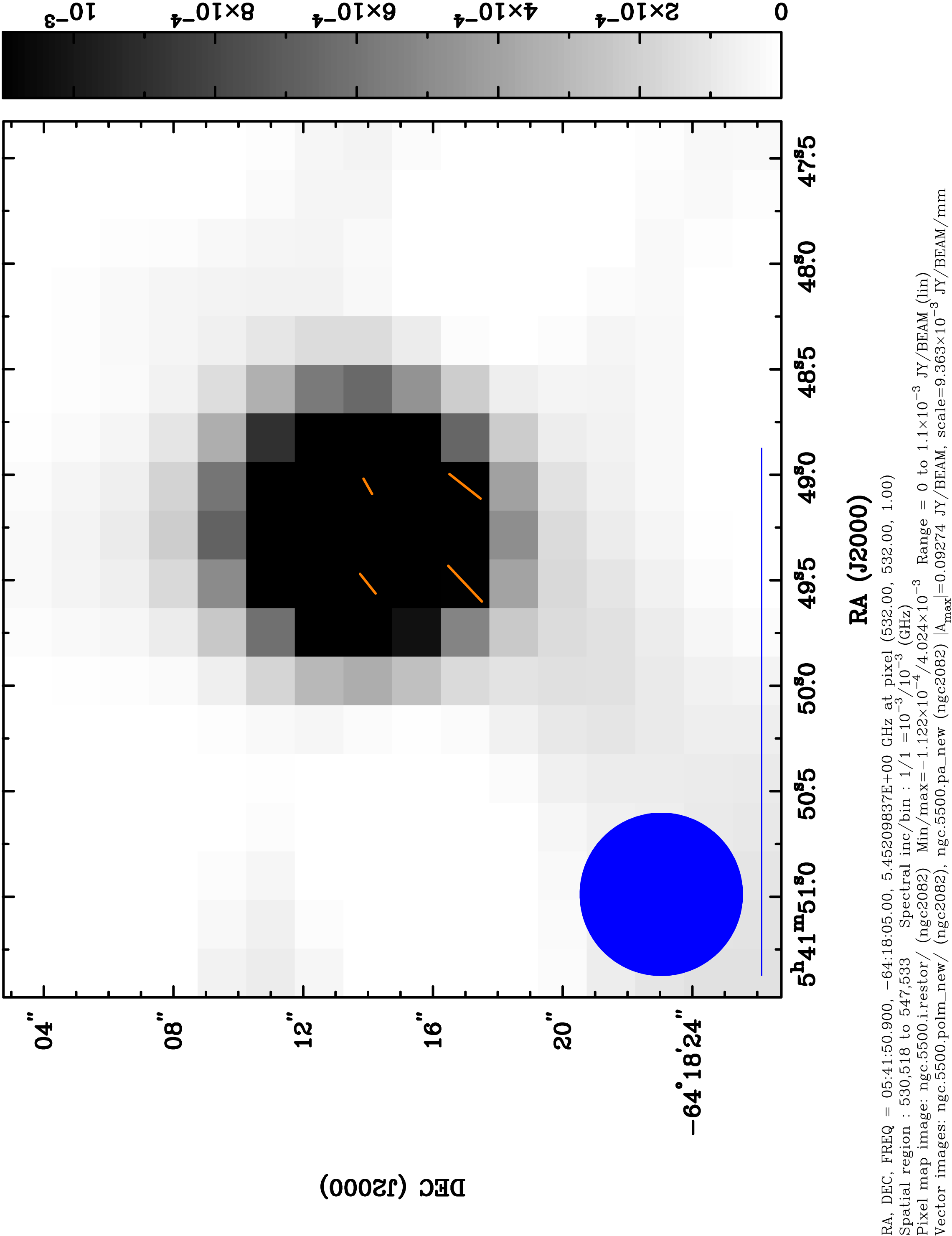}}\quad
\subfloat[Polarisation intensity map of \ngc{} at 5500\,MHz, with contours of 0.0003 and 0.002 Jy\,beam$^{-1}$. \label{fig:pol_int_5500}]{\includegraphics[angle=270,width=.45\textwidth]{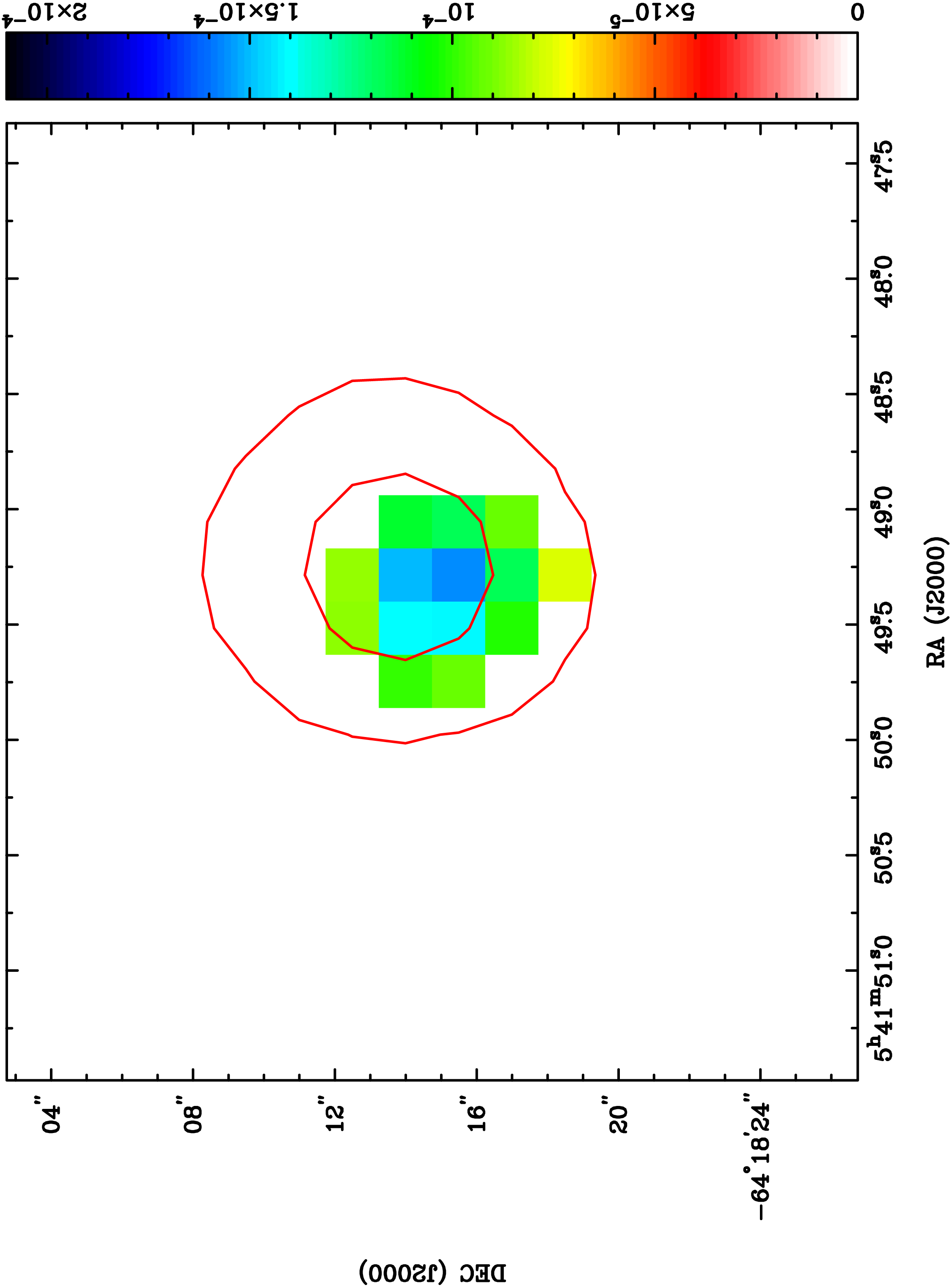}}\\
\subfloat[Fractional polarisation map of \ngc{} at 9000\,MHz, where \textbf{average} $P_{9000\,\text{MHz}} = 8 \pm 4\,\%$. \label{fig:pol_frac_9000}]{\includegraphics[angle=270,width=.45\textwidth, trim=0 20 45 0,clip]{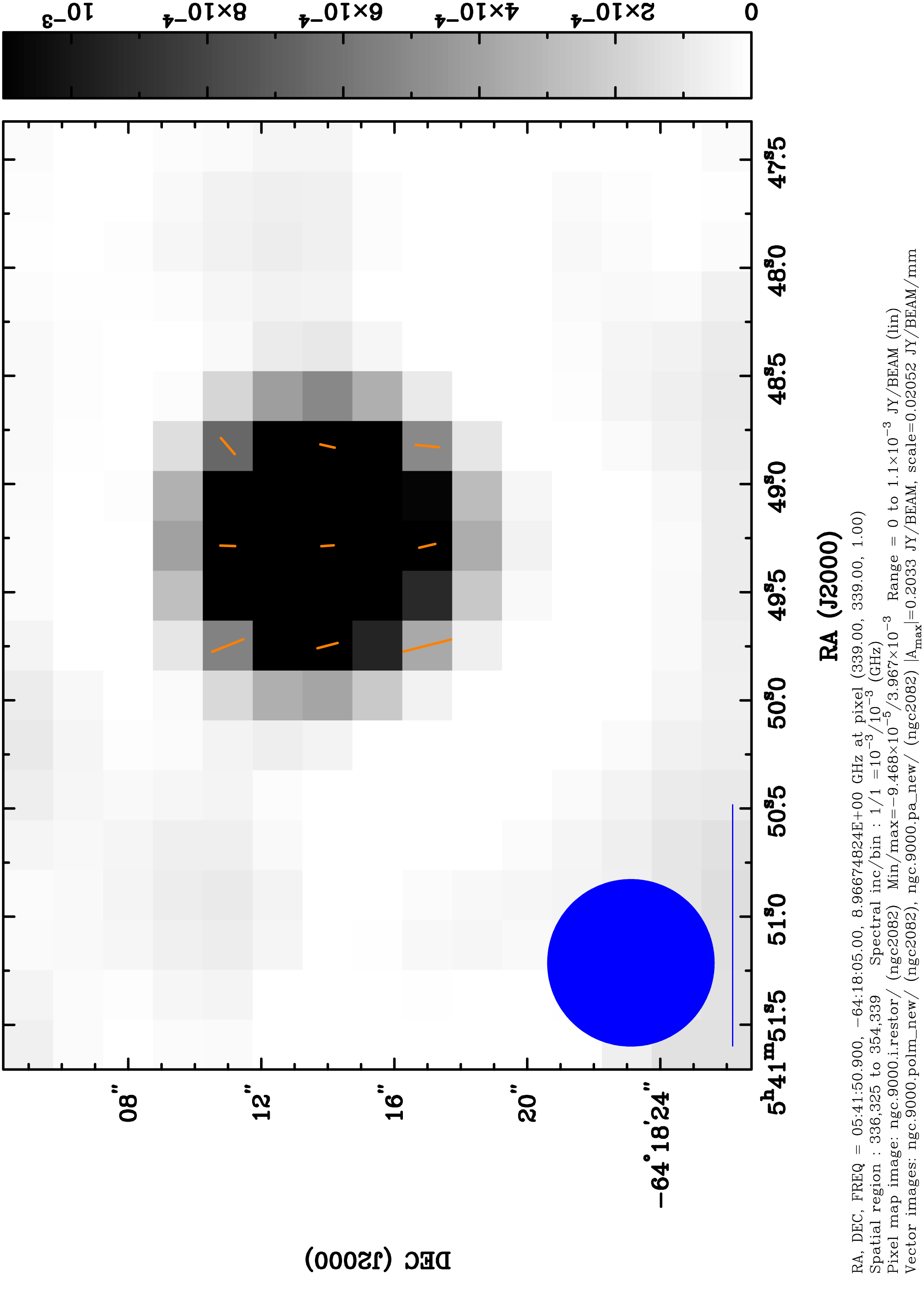}}\quad
\subfloat[Polarisation intensity map of \ngc{} at 9000\,MHz, with contours of 0.0002 and 0.002 Jy\,beam$^{-1}$. \label{fig:pol_int_9000}]{\includegraphics[angle=270,width=.45\textwidth]{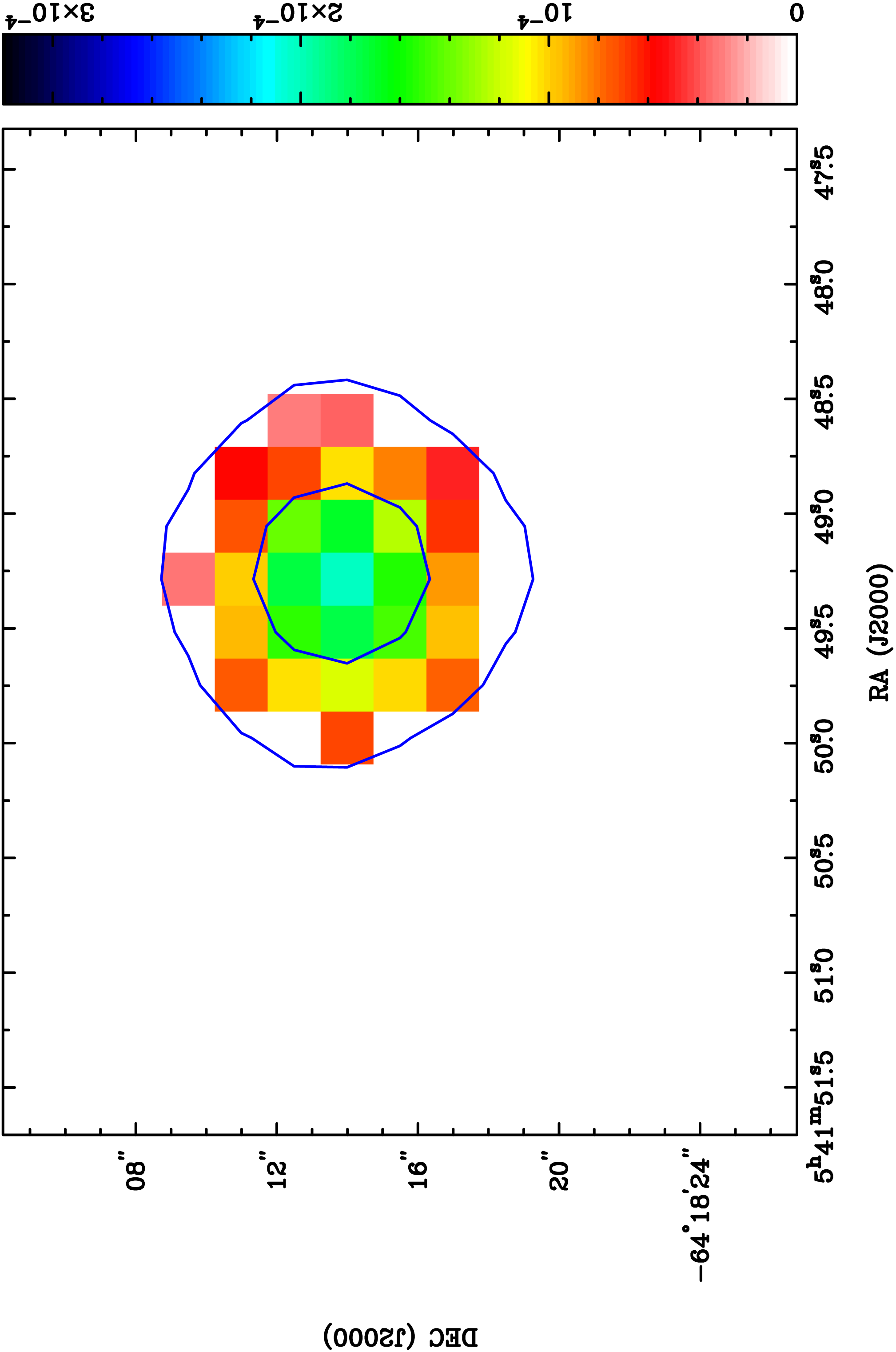}}
\caption{{\bf Left:} Fractional polarisation maps of the radio point source \ngc{}, where the circle on the lower left corner represents a synthesised beam of $5\times5$\,arcsec and the blue line below he circle represents 100\,per\,cent polarisation. The bar on the right side represents the gray-scale gradients for the ATCA image in Jy\,beam$^{-1}$. {\bf Right:} Polarisation intensity maps of \ngc{}, where the bar on the right represents the colour-scale gradients for the ATCA image in Jy\,beam$^{-1}$.}
\end{figure*}

We also investigate if any polarisation from \ngc{} or NGC\,2082 can be detected in our ATCA images. The fractional linear polarisation ($P$) of NGC\,2082 was calculated using the equation:
\begin{equation}
\label{eq: polarisation}
P=\frac{\sqrt{S^2_Q+S^2_U}}{S_I}~,
\end{equation}
\noindent where $P$ is the mean fractional linear polarisation, $S_{Q}$, $S_{U}$, and $S_{I}$ are integrated intensities for the $Q$, $U$, and $I$ Stokes parameters, respectively. We calculate $P_{5500\,\text{MHz}} = 6 \pm 2\,\%$ (see Fig.~\ref{fig:pol_frac_5500}) and $P_{9000\,\text{MHz}} = 8 \pm 4\,\%$ (see Fig.~\ref{fig:pol_frac_9000}). Their associated polarisation intensity maps are seen in Figures \ref{fig:pol_int_5500} and \ref{fig:pol_int_9000} respectively. This weak polarisation associated with \ngc{} is most likely explained if the source is of background origin (see Section~\ref{sec:bckg}).

\subsection{Luminosity of \ngc{}}
 \label{subsec:Luminosity}
 
Assuming a concordance cosmology with parameters provided by \cite{planckcolab}, the luminosity distance of NGC\,2082 is $D_{L} = 17.5\,\text{Mpc}$. 
The radio luminosity of \ngc{} at 888\,MHz is then $L_{888\text{MHz}} = 1.29 \times 10^{20}$\,W\,Hz$^{-1}$ (with $\alpha = 0.02 \pm 0.09$). For comparison, the luminosity of the LMC's SN\,1987A SNR (SNR\,1987A) at its peak flux density ($S_{1000\text{MHz}} = 0.15$\,Jy; \citep{Turtle1987}) is $L_{1000\text{MHz}} = 4.74 \times 10^{16}$\,W\,Hz$^{-1}$ (with $\alpha = -0.74 \pm 0.02$ between 72\,MHz and 8640\,MHz \citep{SN1987aSI}) --- 4 orders of magnitude smaller than \ngc{}. In addition to the very different radio spectral indexes of \ngc{} and SNR\,1987A, which indicate different emission origins, \ngc{} is probably too bright to be an SNR originating from NGC\,2082.

\subsection{An FRB embedded in \ngc{?}}
Fast radio bursts (FRBs) are extremely bright transient events of unknown origin \citep{Lorimer_2007}. While FRBs can be associated with a variety of types of galaxies~\citep[e.g.,][]{bsp+20}, two of the most active repeating FRBs\,121102~\citep{2017Natur.541...58C} and 190520B~\citep{2021arXiv211007418N} are associated with compact persistent radio sources (PRSs). They also show large and variable DM and rotation measure~\citep{msh+18,dfy+22}, indicating extreme magneto-ionic local environment. The compact nature of \ngc{} and its location at the outskirt of NGC\,2082 are reminiscent of those of FRBs\,121102 and 190520B. 

FRB\,121102's PRS (QRS\,121102) has a flat spectrum, with a spectral index of $\alpha = 0.07 \pm 0.03$ \citep{FRB121102SI}, and its luminosity at 1600\,MHz is $L_{1600\text{MHz}} = 2.8 \times 10^{22}$\,W\,Hz$^{-1}$ \citep{law2021fast} --- two orders of magnitude greater than the luminosity of \ngc{}, $L_{1600\text{MHz}} \approx L_{888\text{MHz}} = 1.29 \times 10^{20}$\,W\,Hz$^{-1}$.  \cite{chen2022comprehensive} suggested that QRS121102 is too luminous to be an SNR, consistent with our findings for \ngc{} in Section \ref{subsec:Luminosity}. 

190520Bs PRS (henceforth "QRS\,190520B") has a luminosity $L_{3000\text{MHz}} = 3 \times 10^{22}$\,W\,Hz$^{-1}$ and a spectral index of $\alpha = -0.41 \pm 0.04$ \citep{2021arXiv211007418N}. Similarly, this is 2 orders of magnitude greater than the luminosity of \ngc{}, $L_{3000\text{MHz}} \approx L_{888\text{MHz}} = 1.29 \times 10^{20}$\,W\,Hz$^{-1}$. Unlike QRS\,121102 and \ngc{}, QRS\,190520Bs spectral index is steep and negative --- suggesting non-thermal emission as its origin.

Our Parkes observations (see Section~\ref{subsec:Parkes}) over 3 days detected no transient events. Despite this, considering both FRB\,121102 and 190520B show sporadic outbursts~\citep{rms+20,dfy+22}, it is plausible \ngc{} could host a repeating FRB and we observed during a quiescent period.

\subsection{\ngc{} as a background object?}
\label{sec:bckg}
If intrinsic sources associated with \ngc{} are implausible, the most likely remaining possibility is an extragalactic background source, such as a quasar, radio galaxy or AGN. If so, we may expect to see some \HI\ absorption; however, there is currently no high resolution \HI\ data for NGC\,2082. The flat spectral index together with somewhat weak polarisation at 5500 and 9000~MHz images argue in favour of \ngc{} background origin.

Our observations (Tab.~\ref{tab:flux_density}) show that the flux density at 5500\,MHz is $\sim$4.0\,mJy. From \cite{1994AuJPh..47..625W}, we find that for observations at 5500\,MHz there are $\sim$15\,sources/deg$^2$ at $\geq$4.0\,mJy. The probability of finding a source of such brightness behind NGC\,2082 is then,
\begin{equation*}
    P = A \times 15\,\text{sources/deg}^{2}
\end{equation*}
where $A$ is NGC\,2082s area on the sky in deg$^{2}$. We calculate $P = 1.2\%$, given the radius of NGC\,2082 is $r = 0.016\,\text{deg}$ \citep{deVaucouleurs}.

\section{Conclusions}
 \label{conclusion}
Nearby spiral galaxy NGC\,2082 was found to contain an bright, compact radio source, \ngc{} (Fig.~\ref{fig:contours}), which is most likely of background origin. The flux densities reveal \ngc{} has a flat spectral index, indicating its source may be of thermal origin. We compare the luminosity of \ngc{} to SNR\,1987A, QRS\,121102, and QRS\,190520B, finding that \ngc{} is likely too bright and flat to be a supernova, and is probably not bright enough to be a persistent radio source with an embedded FRB progenitor.

\acknowledgments

The Australia Telescope Compact Array (ATCA) and Australian SKA Pathﬁnder (ASKAP) are part of the Australia Telescope National Facility which is managed by CSIRO. 
Operation of the ASKAP is funded by the Australian Government with support from the National Collaborative Research Infrastructure Strategy. The ASKAP uses the resources of the Pawsey Supercomputing Centre. Establishment of the ASKAP, the Murchison Radio-astronomy Observatory, and the Pawsey Supercomputing Centre are initiatives of the Australian Government, with support from the Government of Western Australia and the Science and Industry Endowment Fund. We acknowledge the Wajarri Yamatji people as the traditional owners of the Observatory site. SD is the recipient of an Australian Research Council Discovery Early Career Award (DE210101738) funded by the Australian Government. This research has made use of the NASA/IPAC Extragalactic Database (NED), which is operated by the Jet Propulsion Laboratory, California Institute of Technology, under contract with the National Aeronautics and Space Administration.
This research made use of APLpy, an open-source plotting package for Python \citep{2012ascl.soft08017R}.
This research is based on observations made with the NASA/ESA Hubble Space Telescope, and obtained from the Hubble Legacy Archive, which is a collaboration between the Space Telescope Science Institute (STScI/NASA), the Space Telescope European Coordinating Facility (ST-ECF/ESA) and the Canadian Astronomy Data Centre (CADC/NRC/CSA).
We thank the anonymous referee for a constructive report and useful comments.

\begin{dataavailability}
All data are publicly available:
\begin{itemize}
    \item Parkes (project code PX075): \\ \tab \url{https://data.csiro.au/}
    \item ATCA CABB (project code C3275) and \\Pre-CABB (project code C466): \\ \url{https://atoa.atnf.csiro.au/query.jsp}
    \item ASKAP (project code AS101): \\ \url{https://data.csiro.au/}
    \item HST (proposal ID 9395): \\ \url{https://hla.stsci.edu/hlaview.html}
\end{itemize}
\end{dataavailability}

\begin{authorcontribution}
Miroslav Filipovi\'c and Shi Dai contributed to the original discovery, conception and design of this study. Parkes data collection and analysis were performed by Joel Balzan and Shi Dai. The first draft of this manuscript was written by Joel Balzan and all authors commented on previous versions of the manuscript. Rami Alsaberi observed and reduced ATCA data. All authors read and approved the final manuscript.
\end{authorcontribution}

\begin{fundinginformation}
No funding was acquired for this research.
\end{fundinginformation}

\begin{ethics}
\begin{conflict}
All authors declare that they have no conflicts of interest.
\end{conflict}
\end{ethics}

\bibliographystyle{spr-mp-nameyear-cnd}
\bibliography{sn-bibliography}

\end{document}